%% file: article.tex
\documentclass[3p,12pt]{elsarticle}

\usepackage{amssymb}
\usepackage{float}
\usepackage{multirow} % for tables
\usepackage{rotating} % for tables (confirmed by mail request that its OK)
\usepackage{verbatim} % for comments 
\usepackage{tabularx}
\usepackage{colortbl}
\usepackage[draft]{todonotes}   % notes showed
\usepackage{enumitem}
\usepackage{lscape} 
 \usepackage{tabularx,booktabs,ragged2e}
\usepackage{tablefootnote}
\usepackage{caption}
\usepackage{booktabs}
\usepackage{soul} % for text highlighting 
\usepackage{longtable}

 \usepackage{makecell}

\journal{Journal of Systems and Software}

\begin{document}

\begin{frontmatter}



\title{Data Pipeline Quality: Influencing Factors, Root Causes of Data-related Issues, and Processing Problem Areas for Developers}
% Influencing Factors and an Empirical Study on Root Causes of Data Pipeline Quality
% Data Pipeline Quality: Influencing Factors and Issues

%Factors and Issues influencing Data Pipeline Quality
%Factors contributing to Data Pipeline Quality:\\A Taxonomy and Empirical Insights}
%Data Pipeline Quality:\\ Influencing Factors and Empirical Insights
%Data Quality in Data Pipelines:\\ Influencing Factors, Root Causes and Challenges
% Factors of Influence on the Data Quality \\ of Data Pipelines
% Factors influencing the Quality of Data delivered by Data Pipelines
% Factors influencing the Data Quality of Data Pipelines
% Data Quality of Data Pipelines: Factors of Influence
% (Data) Pipeline Data Quality: Factors of Influence
% Factors influencing the (Data) Pipeline Data Quality 
% Data Quality of Data Pipelines:\\ Factors of Influence#

% think about replacing "of" with "in" ... "DQ in/of Data Pipelines"


\author[inst1]{Harald Foidl\corref{cor1}}
\ead{harald.foidl@uibk.ac.at}
\cortext[cor1]{Corresponding author}

\affiliation[inst1]{organization={University of Innsbruck},%Department and Organization
           % addressline={Address One}, 
           % city={Innsbruck},
           % postcode={6020}, 
           % state={Austria},
            country={Austria}}

\author[inst1]{Valentina Golendukhina}

  \author[inst2]{Rudolf Ramler}         
            
   \affiliation[inst2]{organization={Software Competence Center Hagenberg GmbH},%Department and Organization
           % addressline={Address One}, 
           % city={Innsbruck},
           % postcode={6020}, 
           % state={Austria},
           %  \department{Software Science}
     	   %  \email{rudolf.ramler@scch.at}
            country={Austria}}
            
\author[inst3,inst1,inst4]{Michael Felderer}

\affiliation[inst3]{organization={Institute for Software Technology, German Aerospace Center (DLR)},%Department and Organization
            %addressline={Address Two}, 
            %city={City Two},
            %postcode={22222}, 
            %state={State Two},
            country={Germany}}            
            
\affiliation[inst4]{organization={University of Cologne},%Department and Organization
            %addressline={Address Two}, 
            %city={City Two},
            %postcode={22222}, 
            %state={State Two},
            country={Germany}}

\begin{abstract}

\input{sections/0-Abstract}

\end{abstract}




\end{frontmatter}



\input{sections/1-Introduction}

\input{sections/2-Background}

\input{sections/3-Research_Procedure}

\input{sections/4-Influencing_Factors}

\input{sections/5-Empirical_Study}

\input{sections/6-Discussion}
\input{sections/7-Conclusions}

\section*{Acknowledgement}

 This work was supported by the Austrian ministries BMK \& BMDW and the State of Upper Austria in the frame of the COMET competence center SCCH [865891], and by the Austrian Research Promotion Agency (FFG) in the frame of the projects Green Door to Door Business Travel [FO999892583] and ConTest [888127]. We also thank Ilona Chochyshvili and Matthias Hauser for their contribution in the conduction of the exploratory case studies.

%\appendix

%\section{Appendix}

 \bibliographystyle{elsarticle-num} 
 \bibliography{literature}





% \end{thebibliography}
\end{document}

%% file: sections/0-Abstract.tex
\noindent 
Data pipelines are an integral part of various modern data-driven systems. However, despite their importance, they are often unreliable and deliver poor-quality data. A critical step toward improving this situation is a solid understanding of the aspects contributing to the quality of data pipelines. Therefore, this article first introduces a taxonomy of 41 factors that influence the ability of data pipelines to provide quality data. The taxonomy is based on a multivocal literature review and validated by eight interviews with experts from the data engineering domain. Data, infrastructure, life cycle management, development \& deployment, and processing were found to be the main influencing themes. Second, we investigate the root causes of data-related issues, their location in data pipelines, and the main topics of data pipeline processing issues for developers by mining GitHub projects and Stack Overflow posts. We found data-related issues to be primarily caused by incorrect data types (33\%), mainly occurring in the data cleaning stage of pipelines (35\%). Data integration and ingestion tasks were found to be the most asked topics of developers, accounting for nearly half (47\%) of all questions. Compatibility issues were found to be a separate problem area in addition to issues corresponding to the usual data pipeline processing areas (i.e., data loading, ingestion, integration, cleaning, and transformation). These findings suggest that future research efforts should focus on analyzing compatibility and data type issues in more depth and assisting developers in data integration and ingestion tasks. The proposed taxonomy is valuable to practitioners in the context of quality assurance activities and fosters future research into data pipeline quality.

%% file: sections/1-Introduction.tex
%%%%%%%%%%%% Introduction %%%%%%%%%%%% 

\section{Introduction} \label{sec:introduction}
Data pipelines play a crucial role in today's data-centric age and have become fundamental components in enterprise IT infrastructures. By collecting, processing, and transferring data, they make it possible to use the ever-growing amounts of information to gain valuable insights and make improved decisions. In addition, they have also become integral parts of data-driven systems, e.g., recommender systems, speech, and image recognition systems. Within these systems, they are primarily responsible for putting the data into a suitable form so that the built-in machine learning (ML) algorithms can automatically make intelligent decisions.

Since the quality of the data plays an important role in making reliable and accurate decisions, research on data cleaning and validation has gained significant interest in the last decade \cite{Breck_etal2019,Biessmann_etal2021}. Most of the work dealt with cleaning and validating data at the early stages of a pipeline, assuming that low data quality was already caused before the pipeline \cite{Foidl&Felderer2019}. The quality of data pipelines themselves, that is, processing data correctly and without errors, has not been treated in much detail. However, the importance of reliable data pipelines was recently underpinned by a global survey of 1,200 organizations. This survey found that 74\% of companies that invest in high-quality and reliable data pipelines have increased their profits by an average of 17\% \cite{IDC_2020}. 

Nevertheless, evidence suggests that data pipelines tend to be error-prone, hard to debug, and require a lot of maintenance and management \cite{Bomanson2019,Romero_etal2020, Munappy_etal2020a}. A recent survey even identified debugging and maintaining data pipelines as the most pressing issues for data engineers \cite{DW&DK_2021}. Moreover, recent studies \cite{Islam_etal2019,Wang_etal2022} show that developers face huge difficulties implementing data processing logic. These difficulties were also observed by Yang et al. \cite{Yang_etal2021}. In their study, they found data handling code to be often repetitive, dense, and error-prone. This bad state of data pipelines contributes to the fact that today's data-driven systems suffer heavily from data-induced bugs and data-related technical debt \cite{Bogner_etal2021,Foidl_etal2022}. 

Research has recently started to address these issues in a variety of ways. First, there are research efforts aiming to improve the debugging of data pipelines \cite{Kossak&Zwick2019,Rezig_etal2020,Lourenco_etal2020}. Second, there are several attempts to automate and guide the creation of data processing components \cite{Bilalli_etal2017,Giovanelli_etal2021,Konstantinou_etal2020}. Third, research appears to be actively shifting its focus toward an end-to-end analysis of data pipelines especially considering the entire inner data handling process \cite{Schaefer_etal2020,Munappy_etal2020b}.

However, this recent research stream is still in its early stages. A deeper understanding of the underlying aspects contributing to successful data pipelines is required to ensure dependable pipelines consistently deliver high-quality data. Such an enhanced insight is instrumental in enabling further research endeavors advancing the quality of data pipelines. 
% factors and circumstances 

This paper aims to address this need as follows. \textit{First}, we aim to identify influencing factors (IFs) that may affect a data pipeline's ability to provide high-quality data. We define an IF or factor of influence as any human, technical, or organizational aspect that may affect the ability of a data pipeline to deliver quality data. Those aspects uncover the core drivers of data pipeline quality and serve as fundamental building blocks in the improvement of data pipeline success. \textit{Second}, we adopt a more technological perspective and seek to understand the nature of data pipeline quality better. In particular, we first look at the root causes and stages of data-related issues in data pipelines by studying GitHub projects. Moreover, we further examine whether there are problem areas for developers that do not correspond to the typical processing stages of pipelines by analyzing Stack Overflow questions. These insights are essential to support debugging activities and to define possible strategies to mitigate data-related issues and can further help uncover specific training needs, focus quality assurance efforts, or discover future research opportunities. The major contributions of the paper are: 

 \begin{itemize}
  \item A \textit{taxonomy} of 41 data pipeline IFs grouped into 14 IF categories covering five main themes: data, development \& deployment, infrastructure, life cycle management, and processing. The taxonomy was validated by eight structured expert interviews.
  \item An \textit{empirical study} about (1) the root causes of data-related issues and their location in data pipelines by examining a sample of 600 issues from 11 GitHub projects, and (2) the main topics developers ask about data pipeline processing by analyzing a sample of 400 Stack Overflow posts. 
\end{itemize}

The remaining paper is structured as follows. First, Section \ref{sec:background_related_work} provides relevant background information on data pipelines and discusses related work. Section \ref{sec:research_procedure} describes the applied research procedure in this paper. Section \ref{sec:general_FOI} presents the developed taxonomy of IFs and its evaluation. Section \ref{sec:empirical_study} elaborates on the root causes of data-related issues and the topics developers face in processing data in the context of data pipelines. Afterward, Section \ref{sec:discussion} discusses the findings and limitations of the study. Finally, the paper is concluded in Section \ref{sec:conclusion}.

%% file: sections/2-Background.tex
\section{Background and related work} \label{sec:background_related_work}
In this section, we first cover background information on data pipelines (Section \ref{subsec:data_pipeline}) and subsequently provide an overview of earlier work related to this paper in Section \ref{subsec:rel_work}.

%--------------------------------------------------------------------
%   Data pipeline
%--------------------------------------------------------------------

\subsection{Data pipeline} \label{subsec:data_pipeline}
This section first presents the concept and architecture of a data pipeline. Afterward, common pipeline types and the underlying software stack of data pipelines are outlined.

\subsubsection{Data pipeline concept} 
The concept of a 'data pipeline' is described differently in the literature, depending on the perspective taken \cite{Agostinelli2021}. Following, we describe a data pipeline first from a theoretical and then from a practical perspective.

Theoretically, a data pipeline refers to a \textit{directed acyclic graph} (DAG) composed of a sequence of nodes \cite{Drocco_etal2017}. These nodes process (e.g., merge, filter) data while the output of one node will be the input of the next node. At least one source node produces the data at the beginning of the DAG, and at least one sink node finally receives the processed data. Considering pipelines from this perspective is often done in mathematical settings to formally describe and analyze data flows (e.g., node dependencies).

From a practical point of view, data pipelines typically constitute a piece of software that automates the manipulation of data and moves them from diverse source systems to defined destinations \cite{Munappy_etal2020a}. Thus, data pipelines represent digitized data processing workflows based on a set of programmed scripts or simple software tools. Given the increasing importance and complexity of data processing, data pipelines are nowadays even treated as \textit{complete software systems with their own ecosystem} comprising several technologies and software tools \cite{Koivisto2019}. The primary purpose of this set of complex and interrelated software bundles is to enable efficient data processing, transfer, and storage, control all data operations, and orchestrate the entire data flow from source to destination. We will adopt this practical point of view in the remaining paper.

\subsubsection{Data pipeline architecture}
\begin{figure}[!hb]
\centering
 \includegraphics[width=0.95\linewidth,keepaspectratio]{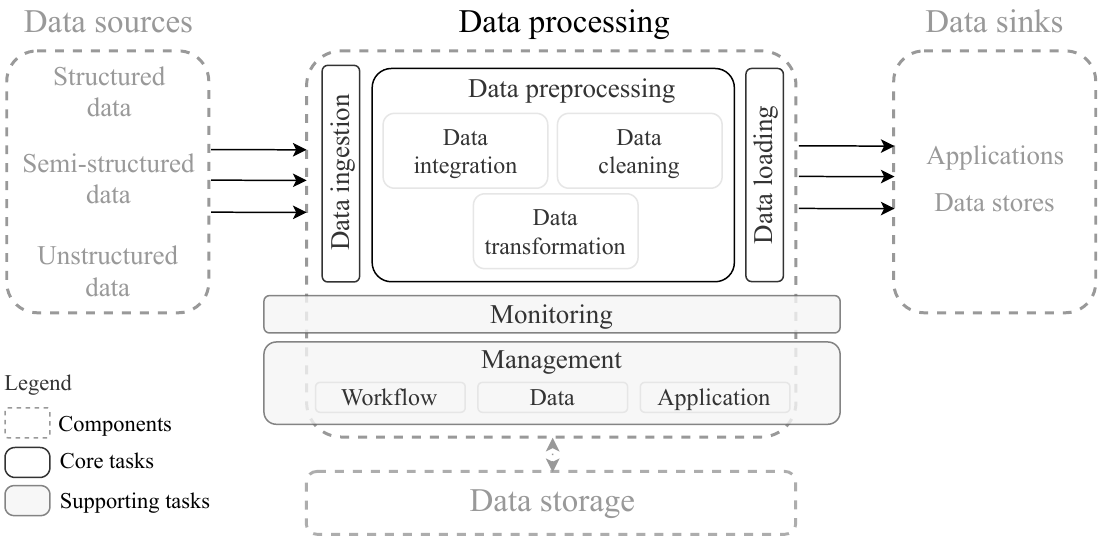}
 \caption{High-level data pipeline architecture
 \label{fig:data_pipe_architecture}}
\end{figure}

We present a generic architecture of a data pipeline shown in Figure \ref{fig:data_pipe_architecture}. As not otherwise stated, the remaining description in this section is based on \cite{Munappy_etal2020a,Hapke&Nelson2020,Munappy_etal2020c,Garcia_etal2016,Chapman2020,Hlupic&Punis2021,Malley_etal2016}.\\

\noindent\textbf{Data pipeline components.}
While the internal structure of a data pipeline can vary significantly, its main components are typically the same: \textit{data sources}, \textit{data processing}, \textit{data storage}, and \textit{data sinks}. Following, we describe these components while focusing on the processing component, as it contains the core tasks of a data pipeline.% (i.e., data ingestion, data preprocessing, and data loading).  

\paragraph{Data sources} 
The starting point of every data pipeline is its data sources. They can be of different types, e.g., databases, sensors, or text files. The data produced by data sources are typically either in a \textit{structured} (e.g., relational database), \textit{semi-structured} (e.g., e-mails, web pages), or \textit{unstructured} (e.g., videos, images) form.

\paragraph{Data processing} 
The second and central component of a data pipeline is its processing component. Within this component, the core tasks for manipulating and handling data (i.e., data ingestion, data preprocessing, and data loading) are provided. We will refer to these core tasks as stages in the remaining paper.

\begin{description}[labelindent=1cm,leftmargin=1cm,font=\normalfont\itshape]%[labelindent=25pt,style=multiline,leftmargin=2.5cm\space]

\item[Data ingestion.]
Typically, the extraction of the data from the data sources and their import in the pipeline starts the data flow. Depending on the size and format of the raw data, different ingestion methods are applied. The data can be ingested into the pipeline with different load frequencies, i.e., the data can be ingested continuously, intermittently, or in batches. 

\item[Data preprocessing.] After the raw data are available in the pipeline, data preprocessing aims to bring the data into an appropriate form for further usage. Typical preprocessing techniques include data integration, cleaning, and transformation. Note that not all preprocessing techniques are applicable in every use case and can be orchestrated differently. \textit{Data integration} aims to merge and integrate the ingested data through, for example, schema mapping or redundancy removal. \textit{Data cleaning} removes data errors and inconsistencies by applying data preparation techniques grouped into missing value treatment (e.g., imputing or discarding) and noise treatment (e.g., smoothing or polishing). \textit{Data transformation} seeks to bring the data in a form suitable for further processing or usage (e.g., statistical analysis or ML models). Typical data preparation techniques for this are binarization, normalization, discretization, dimensionality reduction, numerosity reduction, oversampling, or instance generation.

\item[Data loading.]
This task loads the ingested or preprocessed data into internal storage systems or external destinations. 

\end{description}

\paragraph{Data storage}
The data storage component represents the internal storage of the pipeline. It stores raw, ingested, and processed data, depending on the configuration of the pipeline. A common distinction is made between temporary or long-term storage of the data. 

\paragraph{Data sinks}
The fourth component of a data pipeline describes the destinations where the data are finally provided. These can be other pipelines, applications of any type, or external data storage systems (e.g., data warehouses, databases). \\ \noindent

\noindent\textbf{Data pipeline supporting tasks.}
The data processing tasks of a pipeline are usually supported by several monitoring and management tasks. 

\paragraph{Monitoring} This set of tasks refers to overseeing the data quality, all data operations, and controlling the performance of the pipeline. By using logging mechanisms and creating alerts, monitoring prevents data quality issues and eases debugging and recovering from failures. 

\paragraph{Management} The main management tasks in the context of data pipelines are related to the data, the workflow, and the overall application. Workflow management describes all activities regarding the orchestration and dependencies between the different data processing tasks (i.e., the entire data flow). Data management includes tasks such as creating and maintaining metadata, data catalogs, and data versioning. Application management encompasses the configuration and maintenance (e.g., upgrades, version control) of all software tools a pipeline is built upon.

\subsubsection{Data pipeline types}
Pipelines can be classified based on several characteristics. The three most common types of classification are described in the following. 

\paragraph{Data ingestion strategy}
Data pipelines can be classified based on their data ingestion frequency. Data can be ingested either in batches or continuously. A data pipeline operating in \textit{batch mode} ingests data only in fixed intervals (e.g., daily, weekly) or when a trigger (e.g., manual execution, size of available data) occurs. In contrast, a pipeline continuously ingesting data is operating in \textit{streaming mode}. In this mode, data are consumed in real-time as they become available. There are also scenarios where both ingestion modes are used in parallel (e.g., lambda architecture).

\paragraph{Data processing method}
Another way to classify pipelines is based on the application and order of data processing tasks. Commonly used concepts in this regard are \textit{Extract Transform Load} (ETL) and \textit{Extract Load Transform} (ELT). Originally, ETL was used to describe the data transfer from diverse data sources into data warehouses. Data pipelines applying the ETL concept (i.e., \textit{ETL pipelines}) ingest the data (i.e., Extract), preprocess them (i.e., Transform), and then load (i.e., Load) them to the data sinks. During this usually batch-oriented process, the data are typically stored temporarily in the data storage component of the pipeline. On the other hand, \textit{ELT pipelines} ingest (i.e., Extract) the data and then directly load (i.e., Load) them to the data sinks (e.g., data lakes) without preprocessing them. The preprocessing of the data (i.e., Transform) occurs in the data sinks or by applications consuming these data. As a newer concept compared to ETL, ELT is often applied in cloud-based settings, providing fast data without the need for intermediate storage in the data pipeline.

%ETL: fixed sequences, order crucial, ends with loading
%Pipeline: can comprise anything, order not crucial, must not end with loading

\paragraph{Use case}
Data pipelines can be used for a variety of different purposes. They are typically used for data movement and preparation for, or as part of, other applications (e.g., visualization, analysis tools, ML and deep learning applications, or data mining). A common scenario is the usage of data pipelines to gather data from a variety of sources and to move them to a central place for further usage. In this realm, data pipelines are often referred to as \textit{data collection pipelines} and are applied in nearly every business domain (e.g., manufacturing, medicine, finance). In the context of data science  \cite{Biswas_etal2022}, AI or ML applications, data pipelines are usually used for preparing the data so that they are in a suitable form when fed to algorithmic models. Data pipelines used as part of AI-based systems, ML pipelines, or in data science projects are thus usually referred to as \textit{data preprocessing pipelines}. 

\subsubsection{Data pipeline software stack}
Various technologies and software applications are used for running data pipelines in production. Data processing components can be implemented with different programming languages (e.g., Python, Java), tools or frameworks (e.g., ETL tools such as Apache Mahout or Pig, message brokers such as RabbitMQ or Apache Kafka, or stream processors such as Apache Spark or Flink). Distributed filesystems (e.g., Hadoop Distributed File System) or databases (relational database management systems such as PostgreSQL, or NoSQL such as Apache Cassandra) are usually used to store the data. To coordinate all data processing tasks in a pipeline, workflow orchestration tools are typically used (e.g., Apache Airflow, Apache Luigi). A further group of several tools is used for monitoring the infrastructure, the data lineage, and data quality (e.g., OpenLineage, MobyDQ). From the plethora of tools, one can choose between open-source or proprietary solutions. Moreover, running the pipeline in the cloud is common to address use cases with high demands on scalability.

%--------------------------------------------------------------------
%   Related Work
%--------------------------------------------------------------------

\subsection{Related work}\label{subsec:rel_work}
To the best of our knowledge, no previous study has specifically examined factors that may affect the quality of data in the realm of data pipelines. We thus classify related work into three categories: (1) publications on factors and causes influencing data quality in the field of information and communication technology, (2) contributions on data processing topic issues in related areas, and (3) literature in the context of data pipelines that examine aspects intimately related to the quality of data provided by pipelines.

\subsubsection{Related work on factors and causes that influence data quality}
There is a considerable amount of literature that studies factors influencing data quality. These studies can roughly be classified by the perspective taken and the domain of data being examined. The perspective describes the type of factors (e.g., managerial, technical factors) and the level of abstraction (e.g., very detailed or general factors) considered. The domain of the data describes whether the data represent a specific area (e.g., health, accounting data) or application (e.g., Internet of Things, data warehouse). While many publications treat IFs from a neutral point of view, some treat them from either a positive (e.g., facilitators, drivers of data quality) or a negative (e.g., barriers, impediments of data quality) perspective. 

Several studies \cite{Xu_etal2002, Tee_etal2007,Xiao_etal2009,Xu2013} investigated factors that influence the quality of data in organizations' information systems. Collectively, these studies highlight the importance of management support and communication as crucial factors that influence organizations' general data quality. Other studies took a more narrow approach by focusing, for example, on IFs on master data \cite{Haug_etal2013,Ibrahim_etal2021} or accounting data \cite{Nord_etal2005,Zoto&Tole2014,Xu2015,Knauer_etal2020}. For accounting and master data, literature agrees \cite{Ibrahim_etal2021,Xu2015} that the characteristics of information systems (e.g., ease of use, system stability, and quality) are among the three most influential factors.

In addition to business-related data, there are studies on factors influencing health data quality. Recently, Carvalho et al. \cite{Carvalho_etal2021} identified 105 potential root causes of data quality problems in hospital administrative databases. The authors associated more than a quarter of these causes with underlying personnel factors (e.g., people's knowledge, preferences, education, and culture), thus being the most critical factor for the quality of health data. A different perspective was taken by Cheburet \& Odhiambo-Otieno \cite{Cheburet&Odhiambo2016}. In their study, the authors tried to identify process-related factors influencing the data quality of a health management information system in Kenya. Further, Ancker et al. \cite{Ancker_etal2011} investigated issues of project management data in the context of electronic health records to uncover where those issues arose from (e.g., software flexibility that allowed a single task to be documented in multiple ways). 

With the widespread adoption of the Internet of Things (IoT), there also has been emerging interest in investigating factors that affect IoT data. In 2016, Karkouch et al. \cite{Karkouch_etal2016} proposed several factors that may affect the data quality within the IoT (e.g., environment, resource constraints). A recent literature review of Cho et al. \cite{Cho_etal2021} identified device- and technical-related factors, user-related, and data governance-related factors that affect the data quality of person-generated wearable device data.

The most relevant research regarding our work has investigated factors that influence the quality of data in data warehouses. In 2010, Singh \& Singh \cite{Singh&Singh2010} presented a list of 117 possible causes of data quality issues for each stage of data warehouses (i.e., data sources, data integration \& profiling, ETL, schema modeling). In contrast to Singh \& Singh, a more high-level perspective was taken by Zellal and Zaouia \cite{Zellal&Zaouia2017}. In their work, they examined general factors that influence the quality of data in data warehouses. Therefore, the authors proposed several factors that may affect the data quality loosely based on literature \cite{Zellal&Zaouia2015}. They further developed a measurement model \cite{Zellal&Zaouia2016} to enable the measurement of these factors. Based on this preliminary work, they conducted an empirical study and found that technology factors (i.e., features of ETL and data quality tools, infrastructure performance, and type of load strategy) are the most critical factors that influence data quality in data warehouses. We will compare these contributions with our work in Section \ref{subsec:taxonomy}.

\subsubsection{Related work on data processing topic issues}

Bagherzadeh \& Khatchadourian \cite{Bagherzadeh2019} investigated difficulties for big data developers by mining Stack Overflow questions. They found connection management to be the most difficult topic. In contrast to our work, where we zoom into data processing, they focused on the rather broad and more abstract area of big data. Islam et al. \cite{Islam_etal2019} and  Alshangiti et al. \cite{Alshangiti_etal2019} analyzed the most difficult stages in ML pipelines for developers. Both studies identified data preparation as the second most difficult stage in ML pipelines. Wang et al. \cite{Wang_etal2022} analyzed 1,000 Stack Overflow questions related to the data processing framework Apache Spark. They reported that questions regarding data processing were among the most prevalent issues developers have, with 43\% of all analyzed questions asked. We will compare these contributions with our work in Section \ref{subsec:empirical_studies}.

\subsubsection{Related work on data pipelines}
The quality of the data delivered by data pipelines is closely coupled with the quality of the pipelines themselves. Well-engineered pipelines are likely to provide data of high quality. Following, we give a brief overview of research directions in the context of pipelines that contribute to increasing their quality.

\paragraph{Experiences \& Guidelines}
There is a large volume of literature providing frameworks \cite{Badidi_etal2018,Oleghe&Salonitis_etal2020}, guidelines \cite{Ismail_etal2019,Tardio_etal2020}, or architectures \cite{Ronkainen&Iivari2015,Helu_etal2020} to foster the development and use of data pipelines. While many contributions focused on specific application domains, e.g., manufacturing \cite{O'Donovan_etal2015,Maik&Heinrich2020}, others took a more generic approach \cite{Landesberger_etal2017,Munappy_etal2020b}. Further, there are a number of studies that share experiences (e.g., lessons learned, challenges) about engineering data pipelines \cite{Goodhope_etal2012,Tiezzi_etal2020,Munappy_etal2020a}.

\paragraph{Quality aspects}
Research providing a comprehensive overview of the quality aspects of data pipelines is rare. The studies available mostly focus on specific quality characteristics of pipelines, for example, performance and scalability \cite{Bhandarkar2017,VanDongen&VandenPoel2021} or robustness \cite{Munappy_etal2021}. However, extensive investigations of quality characteristics were conducted on ETL processes \cite{Alves_etal2014,Theodorou_etal2014,Akkaoui_etal2019}. As a sub-concept of data pipelines, ETL processes, and thus their quality characteristics, are highly relevant to data pipelines. 

\paragraph{Development and maintenance support} %Inspection
Recently, several research works have focused on supporting data and software engineers in developing and maintaining data pipelines. There are plenty of works that presented tools for debugging pipelines \cite{Kossak&Zwick2019,Rezig_etal2020,Lourenco_etal2020,Kuchnik_etal2021}. Further, several publications aimed to improve the reliability of pipelines by addressing provenance aspects of pipelines \cite{Wang_etal2019,Rupprecht_etal2020,Grafberger_etal2021}. 

\paragraph{Data Preprocessing}
Besides work on pipelines in general, there is also plenty of research that focuses on the preprocessing of data within pipelines. Studies in this area specifically aim to assist in choosing appropriate data operators and defining their optimal combinations \cite{Bilalli_etal2019,Yan&He2020,Desai&Dinesha2020} or even automate the entire creation of preprocessing chains \cite{Giovanelli_etal2021}.

%% file: sections/3-Research_Procedure.tex
%%%%%%%%%%%% Research Method %%%%%%%%%%%% 

\section{Research procedure} \label{sec:research_procedure}

In this paper, we seek to improve the understanding of data pipeline quality. In fact, this research attempts to address the following objectives.

\begin{enumerate}
   \item To identify and evaluate \textit{influencing factors} of a data pipeline's ability to provide data of high quality.
    \item To analyze \textit{data-related issues} and elaborate on their \textit{root causes} and \textit{location} in data pipelines.
    \item To identify \textit{data pipeline processing topic issues} for developers and analyze whether they correspond to the typical processing stages of pipelines.
\end{enumerate}

\begin{figure}[!h]
\centering
 \includegraphics[width=0.85\linewidth,keepaspectratio]{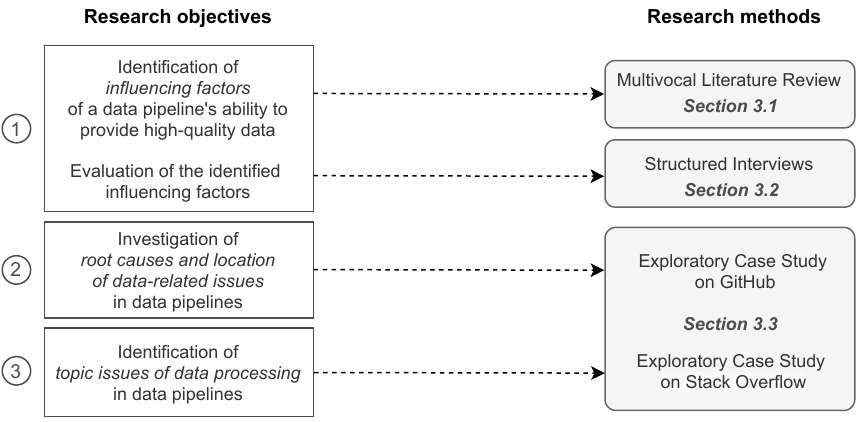}
 \caption{Research procedure
 \label{fig:research_procedure}
}
\end{figure}

An overview of all applied research methods and the corresponding research objectives is depicted in Figure \ref{fig:research_procedure}. The remaining section describes the research methods used to reach these objectives.

%--------------------------------------------------------------------
%   MLR
%--------------------------------------------------------------------

\subsection{Multivocal literature review}
To identify general factors that influence data pipelines regarding their provided data quality, we conducted a Multivocal Literature Review (MLR). The goal of the literature review was to synthesize the available literature related to aspects that may affect the ability of pipelines to deliver high data quality. We gathered problem-related and quality-related aspects of pipelines to derive corresponding IFs.

\subsubsection{Search process and source selection}\label{subsec:searchprocess}
As a first step, we conducted a trial search on the Google and Google Scholar search engines to identify relevant keywords and elaborate the search strategy. While the Google search engine returned a large number of results for the term 'data pipeline', Google Scholar returned instead few in comparison. One possible reason we encountered in our trial search could be that modified versions of the term data pipeline (e.g., data processing pipeline) are often used in the scientific literature, whereas the term data pipeline is commonly used in practice. Another reason could be that research in the developing field of data engineering and its concepts (e.g., data pipeline) is still relatively new. 

\begin{figure}[!h]
\centering
 \includegraphics[width=0.95\linewidth,keepaspectratio]{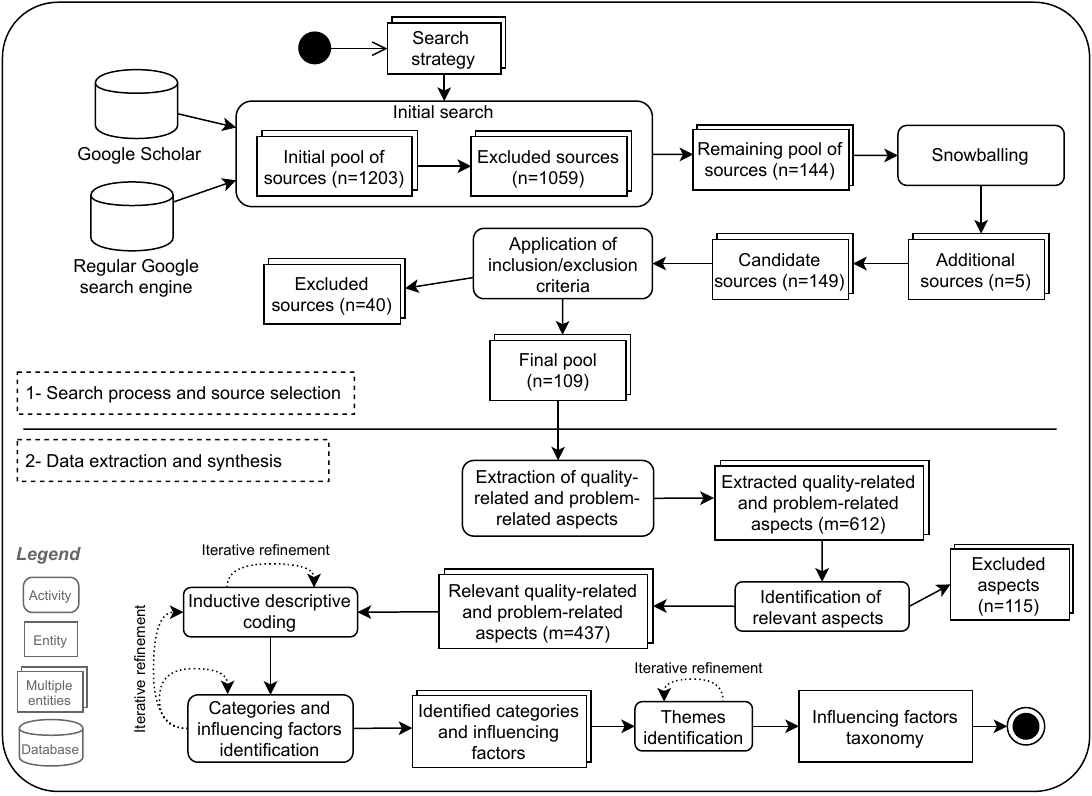}
 \caption{Multivocal literature review process
 \label{fig:MLR}
}
\end{figure}

Because of these differences in the search results, we decided to use different sets of keywords for Google Scholar and the regular Google search engine. The detailed search strings can be found in Table \ref{tab:search_string}. For comprehensibility, the search string used for the regular Google search engine was rewritten with distributive law.

%%In detail, we applied the following search strings (enclosed in square brackets) for Google Scholar: \textit{['data pipeline' AND 'software quality'] OR ['data pipeline' AND 'quality requirements'] OR ['data pipeline' AND 'code quality'] OR ['data pipeline' AND '*functional requirements'] OR ['data pipeline' AND 'quality model'] OR ['data science pipeline'] OR ['data engineering pipeline'] OR ['data pipeline'] OR ['machine learning pipeline'] OR ['data flow pipeline'] OR ['data *processing pipeline'] OR ['data processing pipeline']}. The search strings used for the regular Google search engine were as follows (for comprehensibility rewritten with distributive law): \textit{'data pipeline' AND ('software quality' OR 'quality requirements' OR 'code quality' OR '*functional requirements' OR 'quality model' OR 'pitfall' OR 'error' OR 'anti pattern' OR 'problem' OR 'challenges' OR 'issues')}. 

\begin{table}[!h]
 \caption{Search strings}
 \label{tab:search_string}
 \input{tables/tab-search_strings}
\end{table}

We performed initial searches with these search strings on both search engines. By scanning each hit's title, abstract, and conclusion on Google Scholar, we selected the first set of potentially relevant scientific sources. Potential relevant grey literature was selected based on the title and introduction of each hit on Google's regular search engine.

Based on their generic nature, some keywords (e.g., quality model, error, issue) caused a very large number of hits. To reduce the number of hits to a manageable size, we restricted the search space by utilizing the relevance ranking algorithms of both databases. That is, we assumed that the most relevant sources usually appear on the first few result pages. Thus, we only checked the first three pages of each search string’s result (i.e., 30 hits) and only continued if a relevant source was found on the last page. 

In total, we excluded 1,059 sources from an initial pool of 1,203 scanned sources resulting in a remaining pool of 144 potential relevant sources. Of these, 111 were found with Google Scholar and 33 with Google's regular search engine (i.e., grey literature). 

To ensure finding all relevant sources, we additionally applied forward and backward snowballing \cite{Wohlin2014} to the 111 scientific sources. By examining the references of a paper (backward snowballing) and citations to a paper (forward snowballing), we identified five additional papers. Thus, we got a final set of 149 candidate sources. 

\begin{table}[!h]
\centering
  \caption{Inclusion and exclusion criteria}
  \label{tab:incl_excl_criteria}
\input{tables/tab-incl_excl_criteria}
\end{table}

As a next step, we reviewed each of the 149 sources in detail based on the defined inclusion and exclusion criteria shown in Table \ref{tab:incl_excl_criteria}. To ensure the validity of the results, two researchers independently voted on whether to include or exclude each source. In case of disagreements, the corresponding sources were discussed again, and the final choice was made. Finally, we excluded 40 sources and got a remaining pool of 109 sources (26 grey literature, 83 scientific literature) for further consideration. The complete process of the literature review is depicted in Figure \ref{fig:MLR}.

% -----
% result data (27.05.2022):

% initial search 1203 scanned
% 1059 excluded
% 144 remaining (33 GL & 111 SL 
% +5 snowballing (=SL)
% 149 (33 GL & 116 SL)
% -40 excluded
% 109 final pool (26 GL & 83 SL)

% 612 text fragments
% excluded 175 only high level quality, not data related or too specific 
% 437 coded (164 different codes)
% 104 papers codes
% 5 papers not used because exluded (175 fragments)
% 5 themes, 14 categories, 41 IF
% ----

\subsubsection{Data extraction and synthesis}\label{subsec:dataextraction}
We used a Google Sheets spreadsheet to extract all quality-related (i.e., best practices, quality characteristics) and problem-related (i.e., issues, challenges) aspects of data pipelines from the final pool of sources. In detail, we extracted text fragments describing either quality- or problem-related aspects that may influence the data quality of data pipelines and entered them in separate columns in the spreadsheet. To ensure clarity, we additionally entered context information as notes for some extracted text fragments. Note that whereas some sources contained both quality- and problem-related aspects, others contained information on only one aspect. 

After all text fragments (612) were extracted, we reviewed them based on their potential influence on a data pipeline's ability to deliver high data quality. We excluded text fragments (175) describing either too abstract quality concepts (e.g., the flexibility of pipelines), too general (e.g., data transformation error), or not impacting the quality of data provided by pipelines (e.g., cosmetic bugs of graphical user interfaces). To be able to assess the influence on the data quality, we relied on the ISO/IEC's \cite{ISOIEC2008} inherent data quality characteristics, i.e., accuracy, completeness, consistency, credibility, and currentness.

\begin{figure}[!h]
\centering
 \includegraphics[width=0.95\linewidth,keepaspectratio]{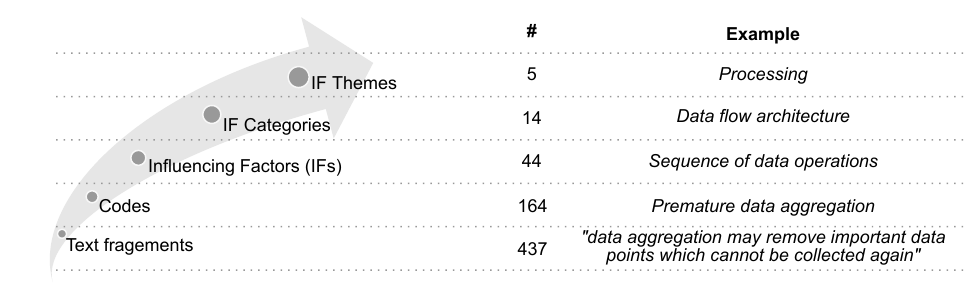}
 \caption{Thematic synthesis procedure
 \label{fig:thematic_synthesis}
}
\end{figure}

The remaining data (437 text fragments) were then synthesized using the thematic synthesis approach. Thematic synthesis is a qualitative data analysis technique that aims to identify and report recurrent themes in the analyzed data \cite{Cruzes&Dyba2011a,Cruzes&Dyba2011b}. This technique was chosen because it is frequently used by studies for similar purposes \cite{Badampudi_etal2016,Fontao_etal2017} and fits well with different types of evidence (i.e., scientific and grey literature) \cite{Cruzes_etal2014}. 

We started the synthesis by applying descriptive labels (i.e., codes) to each data fragment following an inductive approach. Thus, we generated the codes purely based on the concept the text fragment described. In detail, we derived neutral code names that described the underlying influencing aspect of the extracted quality- and problem-related text fragments. After the fragments were coded, two researchers created a visual map of the codes and reviewed them together. Thereby, it was recognized that the level of abstraction varied widely between the codes. Thus, we defined codes that represent higher-level concepts as IF categories (e.g., monitoring) or directly as IFs (i.e., data lineage). In addition, we identified IFs (e.g., code quality) based on the similarity of the codes (e.g., glue code, dead code, duplicated code) and derived overarching IF categories (e.g., software code). During this process, we constantly relabeled and subsumed codes and refined the emerging IFs and categories. In fact, the procedure was carried out in an interactive and iterative manner by two researchers. 

As the last step, the identified IF categories were reviewed together, and a set of concise higher-order IF themes was created to succinctly summarize and encapsulate the categories. The themes were developed based on the experience of the researchers in previous studies \cite{Golendukhina_etal2022,Foidl_etal2022} and refined by reviewing relevant current literature \cite{Lenarduzzi_etal2021,Munappy_etal2020b,Martinez-fernandez_etal2020}. Each researcher then independently assigned all categories to a theme. In the case of different assignments, the corresponding category, as well as themes, were discussed again with a third researcher and refined to reach a consensus. Finally, a terminology control \cite{Usman_etal2017b} of all IFs, categories, and themes was executed to ensure a consistent and accurate nomenclature. Figure \ref{fig:thematic_synthesis} shows the procedure of the thematic synthesis illustrated with an example. In total, we identified five IF themes comprising 14 IF categories and a further 41 IF based on 164 codes assigned to 437 text fragments. 

%--------------------------------------------------------------------
%   Expert Evaluation
%--------------------------------------------------------------------

\subsection{Expert evaluation}
To evaluate our findings and strengthen the trustworthiness of the identified IFs, we conducted structured interviews following the guidelines of Hove \& Anda \cite{Hove&Anda2005} with eight experts to collect empirical evidence about our identified IFs. 

The purpose of the interviews was to validate the identified factors. In total, we collected opinions from eight experts with a minimum of three years of practical experience in the field of data engineering or a similar field where data pipelines are used. 

The interview consisted of two parts: the profiling of the respondents with questions about their experience and background; and the main part with questions about the IFs. To not overwhelm the experts and receive quality feedback, we decided to validate the 14 IF categories and provide the IFs as examples for each category. 

The opinions of the experts regarding the influence of a certain IF category on data quality were measured with a 4-point Likert scale including high, medium, low, or no influence. We also allowed no answer in case participants did not have sufficient experience with a category or were not certain about their answers.

%The categories were then grouped by IF themes with open-ended questions at the end of each theme, asking specifically if the respondents want to share their experience or add something to the description of the IF categories. 

%By synthesising this available knowledge, this paper aims to close this gap.

%--------------------------------------------------------------------
%   Empirical Study
%--------------------------------------------------------------------

\subsection{Empirical study}
\begin{figure} [!h]
    \centering
    \includegraphics[width=1.0\linewidth,keepaspectratio]{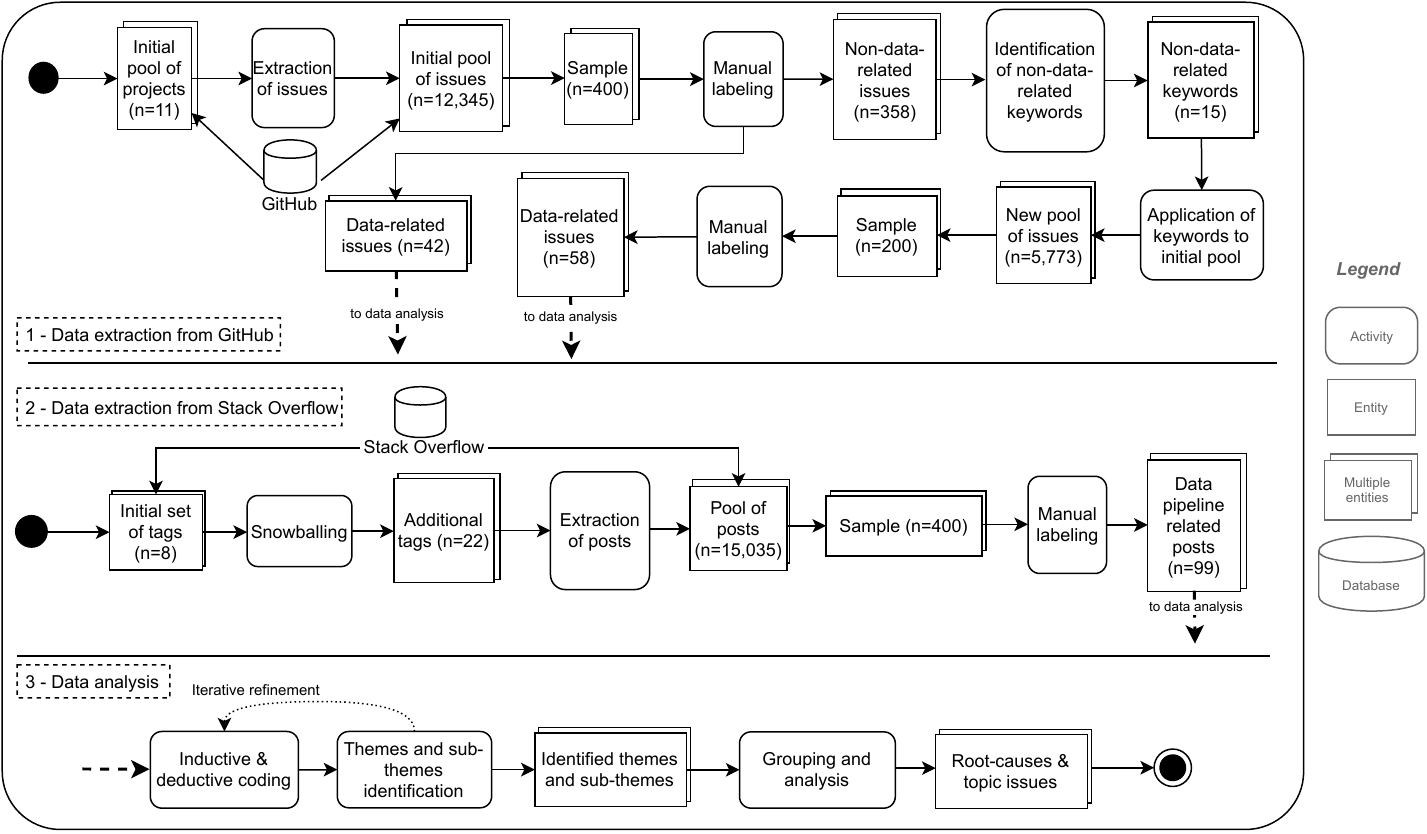}
    \caption{GitHub and Stack Overflow mining procedure}
    \label{fig:research_procedure_emp}
\end{figure}

To get a better understanding of the main problems a data pipeline encounters in its main task, i.e., processing data, we conducted two exploratory case studies. The first study aimed at identifying the root causes of data-related issues and their location in a data pipeline. To achieve this, we analyzed open-source GitHub projects that have a data pipeline as one of its main components. In the second study, we analyzed Stack Overflow posts to identify the main topics developers ask about processing data and examine whether the found problem areas correspond to the typical processing stages of pipelines. The process of mining GitHub and Stack Overflow is presented in Figure \ref{fig:research_procedure_emp} and described in the following sections.

\subsubsection{Exploratory Case Study on GitHub}

\begin{table}[h]
 \caption{Analyzed GitHub projects}
 \label{tab:GitHub projects}
 \input{tables/tab-github_projects}
\end{table}

For mining GitHub, we followed the procedure used by a related study of Ray et al. \cite{Ray2016}. First, we analyzed projects on GitHub and identified those that were suitable for our analysis. The main inclusion criterion was that a project either has a data pipeline as one of the main components or includes several data processing steps before further data application. Following these criteria, we identified 11 projects. The list of the projects, their description, and an overall number of open and closed issues are presented in Table \ref{tab:GitHub projects}.

Next, we extracted 12,345 open and closed issues reported by users or developers from these projects. Since it was unfeasible to analyze over 12,000 posts manually, a random sample of the population was taken considering the confidence level of 95\%. For our population, the minimum sample size is equal to 373. To cover the minimum properly, we chose a sample size of 400. 

In the next step, one researcher manually labeled each issue by assigning three groups: data-related, not data-related, and ambiguous. The last category was added for the issues, including project-specific names or issues that cannot be classified without deeper knowledge of the project. These issues were not included in the further analysis since they do not represent common but rather project-specific issues. As a result, 42 issues were classified as data-related and used in further analysis.

Because only approximately 10\% of all the sample issues were classified as data-related, we decided to apply an additional strategy to increase the number of data-related issues further. The main idea of this strategy was to reduce the total number of extracted issues (12,345) based on a set of keywords that are related to the non-data-related issues. These keywords were extracted from the non-data-related issues from the initial sample. We validated them by checking that they were not present in the data-related issues. The list of the non-data-related keywords is presented in Table \ref{tab:Non-data-related keywords}.

\begin{table}[!h]
\centering
  \caption{Non-data-related keywords}
  \label{tab:Non-data-related keywords}
    \input{tables/tab-non_data_rel_keywords}
\end{table}

Afterward, we applied the set of non-data-related keywords to the total pool of issues and identified 6,172 issues (52\%) as not data-related. In the next step, we took a sample of 200 issues from the remaining 5,773 issues while considering the distribution of all issues in every chosen project, thereby identifying 58 additional data-related issues. As a result, the final set consisted of 100 data-related issues extracted from 11 different projects. 

To assign issues to the root causes and data pipeline’s stages, we applied descriptive labels following an inductive (root causes) and deductive (stages) approach. If the root cause or stage were not identifiable from the issue description, the source code and solution of the issue, if available, were examined. To maintain the objectivity of the results, the labels were then investigated by two other researchers until an agreement was reached.

\subsubsection{Exploratory Case Study on Stack Overflow}
We followed the procedure described by Ajam et al. \cite{Ajam2020} for mining Stack Overflow posts. The process is shown in Figure \ref{fig:research_procedure_emp}.

One of the features of the platform is the ability to add tags to every question according to the topic the given question covers. A tag is a word or a combination of words that expresses the main topics of the question and groups posts into categories and branches. 

If there is more than one topic to which a question belongs, several tags can be applied. Each post can have up to five different tags. To identify the questions relevant to our study, we defined a starting collection of data pipeline processing-related tags. It includes the main tasks of data pipelines defined in the earlier sections such as: 'data pipeline', 'data cleaning', 'data integration', 'data ingestion', 'data transformation', and 'data loading'. In addition, we added two tags, 'Pandas' and 'Scikit-learn', that describe packages frequently used in processing data in pipelines.

To identify and extract posts containing these tags, we used Stack Overflow API, which provides various options for interaction and parsing of the website through commands. API facilitates the process of post-collection and allows the specification of the required tags. Since every post can have up to five tags simultaneously, API helps to extract more related tags based on the initially added tags. Such functionality allows the application of the snowball method where new tags are discovered based on the original ones \cite{Alshangiti_etal2019}. Based on the starting tags, API showed up to eight related tags. We repeated the procedure for several iterations until no new tags were identified. As a result, we got 30 tags shown in Table \ref{tab:Stack Overflow tags} and extracted all posts containing these tags. After handling duplicates, we got a final pool of 15,035 posts from 11,290 different Stack Overflow forum branches.

\begin{table}[!h]
\centering
  \caption{Data pipeline processing-related tags on Stack Overflow}
  \label{tab:Stack Overflow tags}
    \input{tables/tab-stack_overflow_tags.tex}

\end{table}

Since it was unfeasible to analyze over 15,000 posts manually, we randomly selected a sample of 400 posts using a 95\% confidence level. From these posts, we excluded 301 posts asking general questions about the usage of certain libraries or specific data analytics questions (e.g., about image transformation, feature extraction or selection, dimensionality reduction, algorithmic optimization, visualization, or noise reduction).

Afterward, one researcher investigated all 99 remaining sample posts and first assigned deductive labels describing the usual data processing stages of a pipeline. In several iterations, the labels were grouped, refined, and new labels were created until the main data pipeline processing topic issues were identified. The whole labeling process was constantly verified by a second researcher.

%% file: tables/tab-search_strings.tex
\centering
    \begin{small}
    \begin{tabular}{p{0.97\linewidth}}

    \multicolumn{1}{c}{Google Scholar} \\
    \hline
        ['data pipeline' AND 'software quality'] OR ['data pipeline' AND 'quality requirements'] OR ['data pipeline' AND 'code quality'] OR ['data pipeline' AND '*functional requirements'] OR ['data pipeline' AND 'quality model'] OR ['data science pipeline'] OR ['data engineering pipeline'] OR ['data pipeline'] OR ['machine learning pipeline'] OR ['data flow pipeline'] OR ['data *processing pipeline'] OR ['data processing pipeline'] \\  \\

    \multicolumn{1}{c}{Google Search Engine} \\
    \hline
        'data pipeline' AND ('software quality' OR 'quality requirements' OR 'code quality' OR '*functional requirements' OR 'quality model' OR 'pitfall' OR 'error' OR 'anti pattern' OR 'problem' OR 'challenges' OR 'issues') \\
    \end{tabular}
    \end{small}

%% file: tables/tab-incl_excl_criteria.tex
   \begin{small}
   \begin{tabular}{p{0.4\textwidth}p{0.42\textwidth}}
    \toprule
      Inclusion Criteria & Exclusion Criteria \\
   
    \midrule
       Accessible in full-text
        & Non-english articles \\
        
       Published between 2000 and 2021 &
       Only address machine learning aspects \\
        
       Addressing quality characteristics, best \newline  \hspace*{1em} practices, lessons learned, \newline  \hspace*{1em} requirements \textit{or}  problems, issues, \newline  \hspace*{1em} challenges related to data pipelines
       \\

   \bottomrule
  \end{tabular}
  \end{small}

%% file: tables/tab-github_projects.tex
\begin{small}

\begin{tabular}{@{}p{0.03\linewidth}p{0.17\linewidth}p{0.60\linewidth}>{\raggedleft\arraybackslash}p{0.12\linewidth}@{}}
%\begin{tabular}{|c|l|l|c|c|}
\toprule
№ & Project name & Project description & \# of Issues \\
\midrule
1 & ckan & Data management system & 2908 \\ 
2 & covid-19-data & Data collector of COVID-19 cases, deaths, hospitalizations, tests & 888 \\ 
3 & DataGristle & Tools for data analysis, transformation, validation, and movement. & 69 \\ 
4 & dataprep & A tool for data preparation & 376 \\ 
5 & doit & Task management and automation tool & 269 \\  
6 & flyte & Kubernetes-native workflow automation platform for complex, mission-critical data and ML processes at scale & 1442 \\  
7 & networkx & Network Analysis in Python & 2712 \\ 
8 & opendata.cern.ch & Source code for the CERN Open Data portal & 1603 \\ 
9 & pandas-profiling & HTML profiling reports from pandas DataFrame objects & 564 \\
10 & pybossa & A framework to analyze or enrich data that cannot be processed by machines alone. & 999 \\ 
11 & rubrix & Data annotation and monitoring for enterprise NLP & 515 \\ 

\bottomrule
  \end{tabular}
  \end{small}

%% file: tables/tab-non_data_rel_keywords.tex
    \begin{small}
    
    \begin{tabular}{p{0.97\linewidth}}

    \hline
        UI, API, support, version, tutorial, guide, instruction, license, picture, GitHub, typo, logo, documentation, readme, graphics \\
    \hline
    \end{tabular}
    \end{small}

%% file: tables/tab-stack_overflow_tags.tex
\begin{small}
    \begin{tabular}{p{0.97\linewidth}}

    \hline
        arrays, classification, data-augmentation, data-cleaning, dataframe, data-ingestion, data-integration, data-loading, data-pipeline, data-preprocessing, data-science, data-transformation, data-wrangling, datetime, deep-learning, discretization, etl, feature-selection, image-processing, keras, kettle, machine-learning, numpy, opencv, pandas, pdi, pentaho-data-integration , python, scikit-learn, tensorflow
 \\ % random-forest \\
    \hline
    \end{tabular}
\end{small}

%% file: sections/4-Influencing_Factors.tex
%%%%%%%%%%%% General Factors of Influence %%%%%%%%%%%% 

\section{Data pipeline influencing factors} \label{sec:general_FOI}

This section deals with influencing factors (IFs) affecting a pipeline's ability to deliver quality data. First, Section \ref{subsec:general_FOI} presents the IFs in the form of a taxonomy. Afterward, Section \ref{subsec:exp_evaluation} outlines the evaluation of the taxonomy in the form of structured expert interviews.  

%--------------------------------------------------------------------
%   Taxonomy
%--------------------------------------------------------------------

\subsection{Taxonomy of influencing factors}\label{subsec:general_FOI}
\begin{figure}[!hb]
\centering
 \includegraphics[width=0.95\linewidth,keepaspectratio]{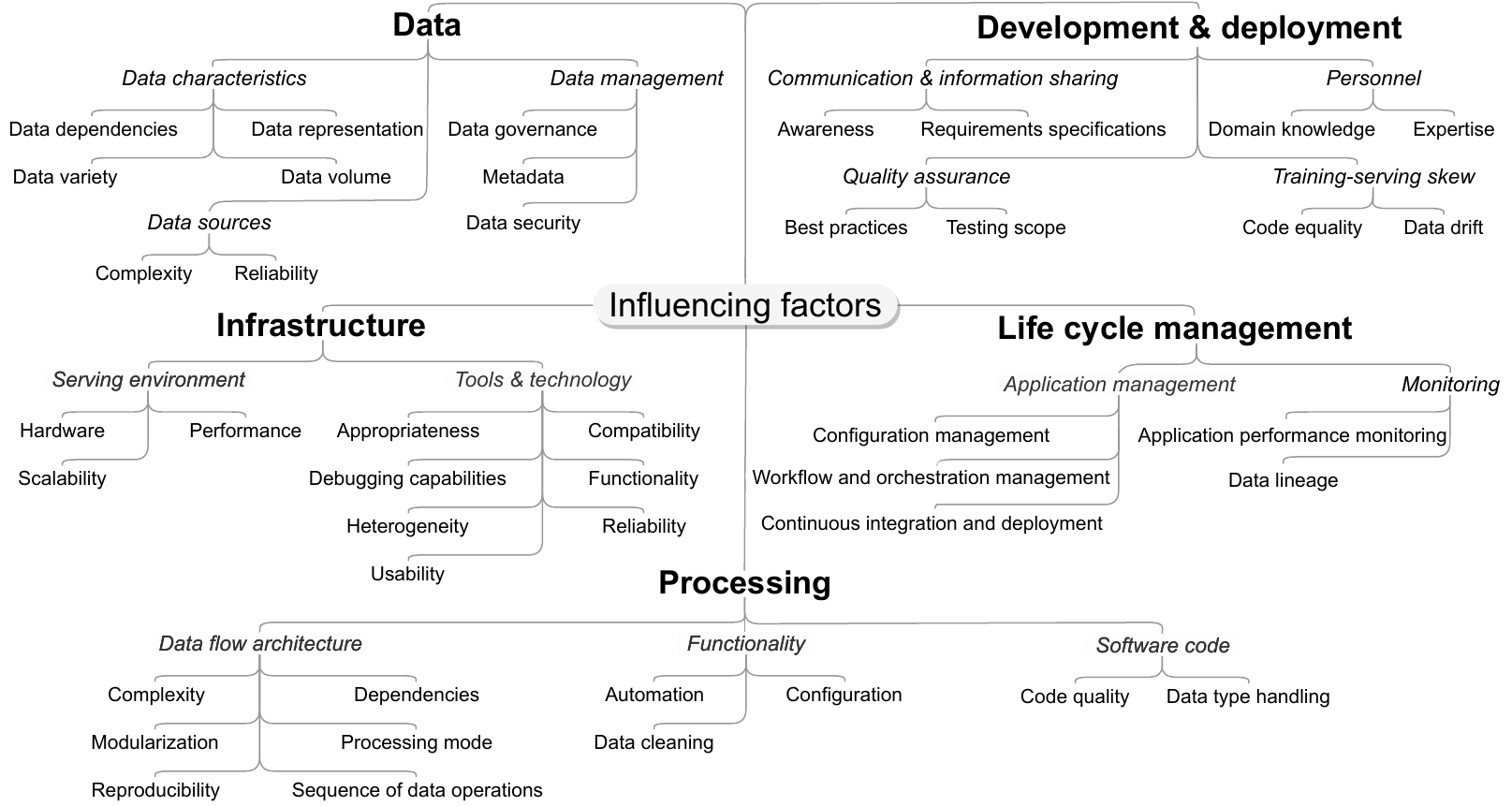}
 \caption{Taxonomy of data pipeline influencing factors (IFs)
 \label{fig:themaic_synthesis}
}
\end{figure}

The taxonomy is depicted in Figure \ref{fig:themaic_synthesis}. In total, we identified 41 IFs grouped into 14 IF categories and five IF themes.  Following, we describe each identified theme, namely \textit{Data}, \textit{Development \& deployment}, \textit{Infrastructure}, \textit{Life cycle management}, and \textit{Processing}, and provide a description of all identified IFs. 

\subsubsection{Data}
This theme covers aspects of the data processed by pipelines that may affect a pipeline's ability to process and deliver these data correctly. The aspects are represented by the following three categories: \textit{Data characteristics}, \textit{Data management}, and \textit{Data sources}. We identified four factors of influence related to the category Data characteristics: 'Data dependencies', 'Data representation', 'Data variety', and 'Data volume'. The category data management comprises the IFs 'Data governance', 'Data security', and 'Metadata'. Note that the IFs data governance and security only relate to the pipeline and not to upstream processes (e.g., data producers). Finally, the 'Complexity' and 'Reliability' of sources providing data are IFs summarized under the category Data sources. A description of each identified IF is given in Table \ref{tab:general_factors_data}.

\begin{table}[h]
 \caption{Influencing factors (IFs) of the theme Data}
 \label{tab:general_factors_data}
 \input{tables/tab-general_factors_data}

\end{table}

\subsubsection{Development \& deployment}
The development and deployment processes of pipelines were identified as significant aspects that may influence a pipeline's data quality. This theme is structured into four categories: \textit{Communication \& information sharing}, \textit{Personnel}, \textit{Quality assurance}, and \textit{Training-serving skew}. Within the category of Communication \& information sharing, we identified the 'Awareness' to perceive a pipeline as an overall construct that comprises different actors and 'Requirements specifications' as IFs. Regarding the category Personnel, 'Expertise' and 'Domain knowledge' were found as important IFs. 'Testing scope' and applying 'Best practices' manifest the IFs of the category quality assurance. Concerning the category Training-serving skew, 'Code equality' between development and deployment and 'Data drift' constitute factors influential to the data quality provided by pipelines. Table \ref{tab:general_factors_dev-dep} provides a description of all identified IFs of this theme.

\begin{table}[h]
 \caption{Influencing factors (IFs) of the theme Development \& deployment}
 \label{tab:general_factors_dev-dep}
 \input{tables/tab-general_factors_dev-dep}

\end{table}

\subsubsection{Infrastructure}
This theme reflects important aspects of the infrastructure data pipelines are based on that may affect their ability to provide data of high quality. The identified IFs of this theme are grouped into two categories: \textit{Serving environment} and \textit{Tools \& technology}. The category Serving environment encompasses infrastructural factors related to pipelines that are running in production, that is, 'Hardware', 'Performance', and 'Scalability'. Under the IF category Tools and technology, we identified the factors 'Appropriateness', 'Compatibility', 'Debugging capabilities', 'Functionality', 'Heterogeneity', 'Reliability', and 'Usability'. This category covers tools and technologies used during development and in the operational state of pipelines. In Table \ref{tab:general_factors_infra} a description of each identified IF of this theme is given.

\begin{table}[h]
 \caption{Influencing factors (IFs) of the theme Infrastructure}
 \label{tab:general_factors_infra}
 \input{tables/tab-general_factors_infra}

\end{table}

\subsubsection{Life cycle management}
The life cycle management of data pipelines has been found to be influential on the quality of the data delivered by pipelines. Within this theme, two IF categories were identified: \textit{Application management} and \textit{Monitoring}. Important IFs of the category Application management are: 'Configuration management', 'Continuous integration and deployment', and 'Workflow and orchestration management'. The category Monitoring comprises the factors 'Application performance monitoring' and 'Data lineage'. Table \ref{tab:general_factors_lcm} describes all IFs of both categories in more detail.

\begin{table}[h]
 \caption{Influencing factors (IFs) of the theme Life cycle management}
 \label{tab:general_factors_lcm}
 \input{tables/tab-general_factors_lcm}

\end{table}

\subsubsection{Processing}

\begin{table}[!h]
 \caption{Influencing factors (IFs) of the theme Processing}
 \label{tab:general_factors_processing}
 \input{tables/tab-general_factors_processing}

\end{table}

This theme includes factors that are related to the processing of the data in pipelines. In total, we identified eleven processing factors that may impact the quality of the processed data. We grouped them into three categories: \textit{Data flow architecture}, \textit{Functionality}, and \textit{Software code}. Regarding the category Data flow architecture, following factors were found to be influential: 'Complexity', 'Dependencies', 'Modularization', 'Processing mode', 'Reproducibility', and 'Sequence of data operations'. The category Functionality summarizes the IFs 'Automation', 'Configuration', and 'Data cleaning'. Finally, 'Code quality' and 'Data type handling' comprise the IFs of the category Software code. Table \ref{tab:general_factors_processing} provides a description of all eleven identified processing factors.

%--------------------------------------------------------------------
%   Expert Evaluation
%-------------------------------------------------------------------

\subsection{Expert evaluation} \label{subsec:exp_evaluation}

To assess the validity of the identified influencing factors, we conducted eight structured interviews with experts from different business areas. A prerequisite for the candidates was at least three years of experience in data engineering. Half of the respondents had three to five years of experience, and the others had more than five years of experience, including two experts with more than ten years of experience. Seven experts assessed all 14 IF categories, while one expert only assessed 11 of them. Additionally, we asked about the format of data being processed in their data pipelines. The main types of data mentioned were tabular data and text. 

\begin{figure} [!h]
    \centering
    \includegraphics[width=1.0\linewidth,keepaspectratio]{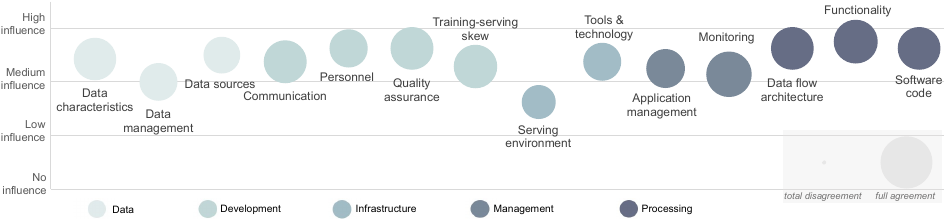}
    \caption{Experts evaluation of IF categories and the levels of their agreement}
    \label{fig:experts_evaluation}
\end{figure}

The results of the experts' evaluation of the taxonomy are presented in Figure \ref{fig:experts_evaluation}. The Y-axis represents the average influence of all 14 assessed IF categories assessed by the experts. Each category is represented by a bubble while the color of the bubble highlights the corresponding IF theme. The size (i.e., area) of the bubbles represents the experts' level of agreement on every IF category and is calculated as the standard deviation. High standard deviation coefficients evidence low agreement among the experts, which is expressed by smaller sizes of the bubbles. As a reference, the two gray shaded bubbles on the lower right-hand side of the figure visualize the range between no and full agreement. 

%We did not aim at or expect a high inter-rater reliability, because of the different experts’ backgrounds in combination with leaving the concrete envisioned use cases open for individual interpretation. Instead, we

The majority of categories, eight out of 14, were assessed similarly by experts, i.e., within one point (influence level) difference. The largest disagreement appeared in five categories: data management, data sources, personnel, serving environment, and tools \& technology. Although the average assessed influence of serving environment-related IFs received the lowest score, it also showed the lowest agreement among experts from no influence to medium influence. Notably, this category received a "no influence" assessment from the expert who did not assess all categories. Besides the three no answers, there were 51 high, 49 medium, eight low, and one no influence assessments.

Despite different levels of agreement, 13 out of 14 categories were identified by the majority of experts to have medium to high influence. According to the experts' assessment, the five most influential categories are functionality, personnel, quality assurance, data flow architecture, and software code. They fall under the development and deployment, and processing categories. Overall, processing-related IFs were rated highly influential, with a high level of agreement among the experts. In conclusion, the expert interviews have confirmed that the categories have the potential to affect the data quality of a data pipeline.

%were noted. they are categorized as ...

% to understand the interplay of factors which

%these factors usually consider the experience and capabilities of

%Project factors regard various qualities of project m

%this factor may facilitate the improvement of the

%were apparent from the data ...

%\paragraph{Communication \& information sharing}
%with the most notable being the ... 

%Product factors cover the characteristics of
%Personnel factors reflect the characteristics of personnel involved in the %software development project

%Process factors are concerned with the characteristics

%The most commonly identified factors represent complex phenomena

%Regarding software domain-specific factors, there are several significant differences 

%The productivity factors observed in the most recent IESE studies d

%The studies presented in the literature provide evidence not only that factors influencing productivity vary across different application domains, b

%Technology capability refers to ability of IT infrastructures

%\begin{table}[h]

% \caption{Inclusion and exclusion criteria.}
%  \label{tab:incl_excl_criteria}
%    \input{tables/tab-general_factors}
%\end{table}

%\input{tables/tab-general-factors_2}

%% file: tables/tab-general_factors_data.tex
\begin{small}

\begin{tabular} {@{}p{0.05\textwidth}p{0.23\textwidth}@{}p{0.72\textwidth}@{}}
\toprule
\multicolumn{2}{l}{IF Category \& IF} & Description \\
\midrule
    
\multicolumn{2}{l}{\textit{Data characteristics}} & \\
& Data dependencies & Instability of data regarding their quality or quantity over time.  \\
& Data representation & Data formats, data structures, data types, and further data representation aspects (e.g., data encoding). \\
& Data variety & Heterogeneity (i.e., formats, structures, types) of the data consumed.  \\
& Data volume & Quantity and rate at which data are coming into the pipeline. \\ \addlinespace

\multicolumn{2}{l}{\textit{Data management}}     &    \\
& Data governance & Processes, policies, standards, and roles related to managing the data processed by a pipeline (e.g., data ownership, data accessibility, raw data storage). \\
& Data security & Degree to which data are protected during their transmission and processing. \\
& Metadata & Degree to which data consumed and processed are documented (e.g., data catalog, data dictionary, data models, schemas). \\  \addlinespace

\multicolumn{2}{l}{\textit{Data sources}}    &        \\
& Complexity & Number of data models and data sources a pipeline has to deal with, including the degree of simplicity of merging the corresponding data (e.g., joinability). \\
& Reliability & Degree to which data sources are available, accessible, and provide data of high quality. \\

\bottomrule
\end{tabular}
\end{small}

%% file: tables/tab-general_factors_dev-dep.tex
\begin{small}

\begin{tabular} {@{}p{0.05\textwidth}p{0.21\textwidth}@{}p{0.74\textwidth}@{}}
\toprule
\multicolumn{2}{l}{IF Category \& IF} & Description \\
\midrule
    
\multicolumn{3}{l}{\textit{Communication \& information sharing}}  \\
& Awareness & The perception of the pipeline as a coherent overall construct and the awareness and ability of knowledge exchange.  \\
& Requirements \vspace{-0.1cm} \newline \hspace*{1em} specifications & Completeness and level of detail regarding the specification and documentation of the pipeline (e.g., data transformation rules, processing requirements).  \\ \addlinespace

\multicolumn{2}{l}{\textit{Personnel}}         \\
& Domain knowledge & Entities’ knowledge and understanding of the domain underlying the data. \\
& Expertise & Background, experiences, technical knowledge, and quality awareness of entities.\vspace*{0.2 cm}  \\ \addlinespace

\multicolumn{2}{l}{\textit{Quality assurance}}            \\
& Best practices & Code and configuration reviews, refactoring, canary processes, and further best practices (e.g., use of control variables). \\
& Testing scope & Testing depth and space (e.g., test coverage, test cases).  \\ \addlinespace

\multicolumn{2}{l}{\textit{Training-serving skew}}        \\
& Code equality & Equality of software code between development and production (e.g., ported code, code paths, bugs). \\
& Data drift & Differences in the data between development and production (e.g., encodings, distribution). \\

\bottomrule
\end{tabular}
\end{small}

%% file: tables/tab-general_factors_infra.tex
\begin{small}

\begin{tabular} {@{}p{0.05\textwidth}p{0.21\textwidth}@{}p{0.74\textwidth}@{}}
\toprule
\multicolumn{2}{l}{IF Category \& IF} & Description \\
\midrule
    
\multicolumn{3}{l}{\textit{Serving environment}}  \\
& Hardware & Type and reliability of the hardware in production. \\
& Performance & Availability and manageability of the resources in production. \\
& Scalability & The capacity to change resources in size or scale.   \\ \addlinespace

\multicolumn{3}{l}{\textit{{Tools \& technology}}}         \\
&  Appropriateness & Types (e.g., code/GUI-first, own solutions, propriety), maturity, flexibility, and up-to-dateness of tools and technology. \\
&  Compatibility & Ability of tools and technology (e.g., frameworks, platforms, libraries) to work together. \\
&  Debugging \vspace{-0.1cm} \newline \hspace*{1em} capabilities & Availability and extent (i.e., level of detail) of opportunities (e.g., tools) to find and correct errors. \\
&  Functionality & The scope of functions (e.g., advanced data operations, support of data management, engineering, and validation) provided. \\
&  Heterogeneity & Number of different technologies and tools used.  \\
&  Reliability & Degree to which tools and technologies are working correct (e.g., software quality, complexity). \\
&  Usability & Availability, documentation, and ease of use of APIs, libraries, and tools. \\

\bottomrule
\end{tabular}
\end{small}

%% file: tables/tab-general_factors_lcm.tex
\begin{small}

\begin{tabular} {@{}p{0.05\textwidth}p{0.35\textwidth}@{}p{0.6\textwidth}@{}}
\toprule
\multicolumn{2}{l}{IF Category \& IF} & Description \\
\midrule
    
\multicolumn{3}{l}{\textit{Application management}}  \\
&  Configuration management & Version control and dependency management for code, libraries, and configurations. \\
&  Continuous integration and \vspace{-0.1cm} \newline \hspace*{1em}  deployment & Automation of providing new releases and changes.\\ 
&  Workflow and orchestration \vspace{-0.1cm} \newline \hspace*{1em} management & Tools for managing the entire workflow of pipelines.  \\ \addlinespace

\multicolumn{3}{l}{\textit{{Monitoring}}}         \\
& Application performance \vspace{-0.1cm} \newline \hspace*{1em} monitoring  & Observing and logging the operational runtime of all software components. \\
& Data lineage & Observability of the data during all transformation steps, including data versioning. \\

\bottomrule
\end{tabular}
\end{small}

%% file: tables/tab-general_factors_processing.tex
\begin{small}

\begin{tabular} {@{}p{0.05\textwidth}p{0.23\textwidth}@{}p{0.72\textwidth}@{}}
\toprule
\multicolumn{2}{l}{IF Category \& IF} & Description \\
\midrule
    
\multicolumn{3}{l}{\textit{Data flow architecture}}  \\
&  Complexity & Computational effort and pipeline complexity (e.g., transformation complexity). \\
&  Dependencies & Dependencies between architectural components or processing steps. \\
&  Modularization & Modularization of data pipeline components (e.g., codes, APIs). \\
&  Processing mode & Type of processing (e.g., batch, stream, distributed or parallel processing). \\
&  Reproducibility & Degree to which processing is reproducible, including aspects such as caching and idempotency. \\
&  Sequence of data\vspace{-0.1cm} \newline \hspace*{1em}  operations & Application order of data preparation and processing techniques.  \\ \addlinespace

\multicolumn{3}{l}{\textit{Functionality}}         \\
&  Automation & Automation of data validation, handling, and providing corresponding guidance. \\
&  Configuration & Configuration (e.g., parameters, settings) for processing data. \\
&  Data cleaning & Treatment of data issues (e.g., modification or deletion).  \\ \addlinespace

\multicolumn{3}{l}{\textit{Software code}}  \\
&  Code quality & Quality and complexity of the code for processing data (e.g., dead code, duplicated code, language heterogeneity). \\
&  Data type handling & Handling of data types (e.g., type casting, delimiter handling, encoding).  \\

\bottomrule
\end{tabular}
\end{small}

%% file: sections/5-Empirical_Study.tex
\section{Empirical study on data-related and data processing issues} \label{sec:empirical_study}

In this section, we first outline the root causes of data-related issues and their location in data pipelines based on analyzing GitHub projects (Section \ref{subsec:GitHub_results}). Afterward, Section \ref{subsec:SO_results} presents the main data pipeline processing problem areas for developers identified by mining Stack Overflow posts and compares them to the typical processing stages of pipelines.

%--------------------------------------------------------------------
%   Root causes and stages of data-related issues in data pipelines (GitHub)
%--------------------------------------------------------------------

\subsection{Root causes and stages of data-related issues in data pipelines}
\label{subsec:GitHub_results}

\begin{figure}[!h]
\centering
 \includegraphics[width=0.95\linewidth,keepaspectratio]{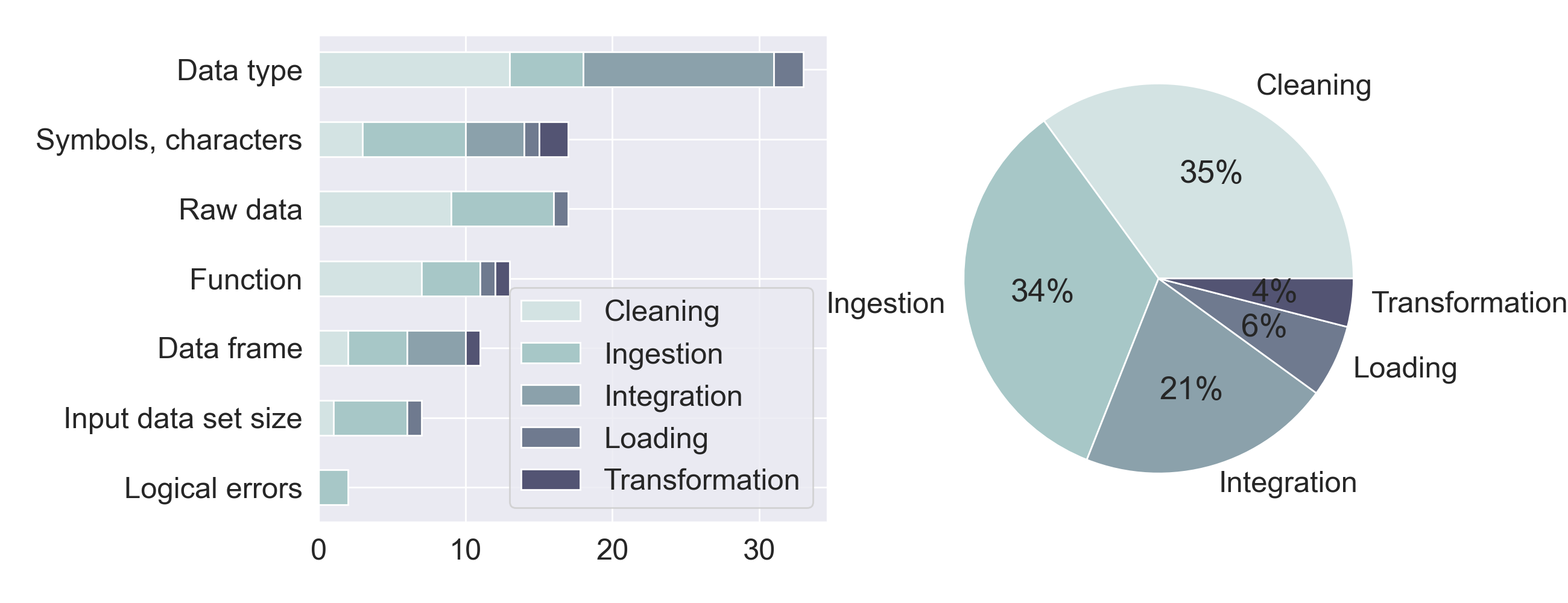}
 \caption{Data-related issues' root causes and pipeline stage in GitHub projects
 \label{fig:fig-root_causes}
}
\end{figure}

After manually analyzing 100 data-related issues from the chosen GitHub projects, we identified seven root cause categories of these issues. The categories are listed on the left-hand side of Figure \ref{fig:fig-root_causes}. 

The most frequent root cause identified in 33\% of the analyzed problems is related to \textit{data types}. These issues occur at almost every stage of the data pipeline; incorrectly defined data types make it difficult to clean, ingest, integrate, process, and load the data. It can lead to more obvious issues when the data cannot be processed due to unknown or incorrect data types, but also it can cause loss of information if data cannot be read and no error is raised. Data type issues are not limited to single data items but also concern how data types are handled in data frames. In almost 90\% of all cases, data type issues arise in the cleaning or integration stages.

Issues caused by the misplacement of \textit{symbols and characters} account for 17\% of all investigated issues. Special characters, such as diacritics, letters of different alphabets, or symbols not supported by used encoding standards, cannot be processed and cause errors in the data pipeline. 

The next identified category of root causes is related to \textit{raw data}. This category describes all issues rooted in the raw data. For example, duplicated data, missing values, and corrupted data. Mostly, the issues appear at the ingestion and integration stages. 

\textit{Functionality} issues account for 13\% of the issues and describe misbehavior of functions or lack of necessary functions. About half of the issues occurred during the cleaning process and characterize wrong outputs of the cleaning functions, i.e., correctly processed data are recognized as incorrect after the cleaning stage of a pipeline. Further issues describe functions delivering inconsistent results or not processing the data as the function intended it.

\textit{Data frame-related} issues were identified in 11\% of all issues. They cover all data frame-related activities, such as data frame creation, merging, purging, and other changes. Additionally, some of the issues are connected to the access of different groups of users to the data.

The last two categories were related to processing large data sets and logical errors, with seven and two percent, respectively. The \textit{input data set size} can cause problems at different pipeline stages. The issues discovered during the analysis included failure to read, upload, and load large data sets. Therefore, the ability of the data pipeline to scale according to the amount of data must be considered already in the early stages of data pipeline development. Issues in the category \textit{logical errors} describe, for example, calls to the non-existing attributes of objects or non-existing methods.

An overview of the pipeline stages where the analyzed data-related issues occurred is shown on the right-hand side of Figure \ref{fig:fig-root_causes}. Most of all issues manifested during the cleaning and ingestion stages of the pipelines, with 35\% and 34\%, respectively. Further, 21\% of all issues were detected at the integration stage. The stages with the fewest issues seen are loading and transformation.

%--------------------------------------------------------------------
%   Main issues of data processing in data pipelines (Stack Overflow)
%--------------------------------------------------------------------

\subsection{Main topic issues of data processing for developers in data pipelines}
\label{subsec:SO_results}

\begin{figure}[!h]
\centering
 \includegraphics[width=0.95\linewidth,keepaspectratio]{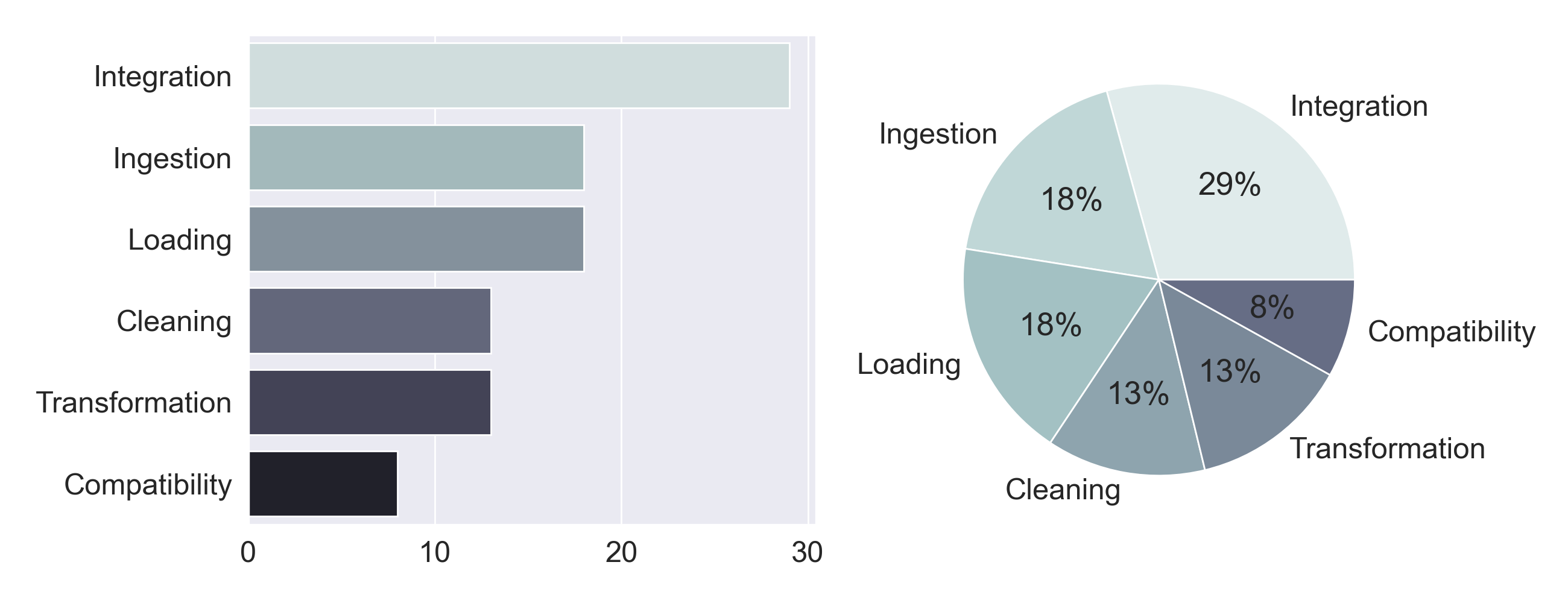}
 \caption{Data pipeline processing topics asked in Stack Overflow posts
 \label{fig:stages_stackoverflow}
}
\end{figure}

After manually labeling all 99 data pipeline processing-related posts, we identified six main data pipeline processing topic issues, five of which represent typical data pipeline stages: integration, ingestion, loading, cleaning, and transformation. The sixth topic identified was compatibility. Figure \ref{fig:stages_stackoverflow} shows the number of posts related to different data pipeline stages and their distribution. Since one post may include more than one label, there are more than 99 labels represented in the figure.

The largest topic was related to \textit{integration} issues. Here users mostly asked questions about the transformation of databases, operations with tables or data frames, and platform- or language-specific questions. 

Posts related to ingestion and loading were the second and third most asked group of posts. Posts related to \textit{ingestion} typically discussed the upload of certain data types and the connection of different databases with data processing platforms. Posts regarding \textit{loading} mostly cover several issues: efficiency, process, and correctness. Efficiency relates to the speed of loading and memory usage. Posts about processes ask platform-related questions on how to connect different services. Correctness covers questions about incorrect or inconsistent results of data output.

Cleaning and transformation posts account for 13\% of all posts. A large portion of all posts regarding \textit{cleaning} include questions about the handling of missing values. Posts that were labeled as \textit{transformation} cover such issues as the replacement of characters and symbols, data types handling, and others. 

The last topic identified is \textit{compatibility}. This topic is not specific to any data pipeline stage but includes a block of questions regarding software or hardware compatibility for data pipeline-related processes. Examples are code running inconsistently in different operational systems, e.g., Ubuntu and macOS, programming language differences, software installation in different environments, and others. Compatibility issues are critical but hard to foresee. We found 8\% of posts on Stack Overflow with a focus on compatibility issues in various stages of data pipelines. In all cases, the issues affected the core functionality of pipelines.

%% file: sections/6-Discussion.tex
\section{Discussion} \label{sec:discussion}

This section discusses the findings of the research in connection with the related literature and describes the limitations of the study. Section \ref{subsec:taxonomy} focuses on the taxonomy and its evaluation. Afterward, Section \ref{subsec:empirical_studies} highlights the core findings and limitations of the exploratory case studies on GitHub and Stack Overflow and connects them to related works. Finally, Section \ref{summary} provides an overview of the findings and their relations.

%--------------------------------------------------------------------
%   Data pipeline influencing factors taxonomy
%--------------------------------------------------------------------

\subsection{Data pipeline influencing factors taxonomy} \label{subsec:taxonomy}
Our developed taxonomy summarizes the factors influencing a data pipeline's ability to deliver high-quality data into five main pillars: data, development and deployment, infrastructure, life cycle management, and processing.

\paragraph{Interpretation}
The identification of 41 IFs grouped into 14 categories underlines the complex and wide range of aspects that may affect data pipelines. This necessitates an urgent need to further study the influencing effects of each factor. Notwithstanding the relatively limited size of our evaluation, the following conclusions can be drawn. Processing-related IFs were identified as highly influential by the majority of the experts in the structured interviews. In contrast, IFs of the theme infrastructure were estimated to have low to medium influence with, however, low agreement rates between respondents. This leads to several conclusions. First, the impact of different IFs on the final data quality is not equal and homogeneous, and some IFs have a higher effect than others. Consequently, they should be considered in the first place. Second, the lower levels of experts' agreement regarding data sources, personnel, and the serving environment might indicate that the influence of different factors depends on the domain of the data pipeline application and the form of data (e.g., tabular, text, images). However, since the main purpose of the evaluation was to confirm the influencing ability of the factors, the results in terms of the relative influencing ranking must be interpreted with caution.

\paragraph{Comparison with related work}
We briefly compare our taxonomy to two contributions closest to our work regarding the domain of data and the perspective taken. The first article was published by Singh \& Singh \cite{Singh&Singh2010} and presents possible causes of data quality issues in data warehouses. In contrast to our work, the authors only consider the causes of poor data quality at a very detailed level and do not form overarching causes (i.e., IFs categories and themes). Moreover, they did not provide empirical validation of the proposed factors. The second article \cite{Zellal&Zaouia2017} took a more high-level perspective and investigated general factors that influence the quality of data in data warehouses. In their work, Zellal \& Zaouia found that technological factors (e.g., ETL and data quality tools), data source quality, teamwork, and the adoption of data quality management practices are the most critical factors that influence data quality in data warehouses. As these factors are also represented in our taxonomy (i.e., tools \& technology, data sources, communication \& information sharing, and best practices), our findings confirm their influential characteristics. Unlike Zellal \& Zaouia's work, however, we base our research on a comprehensive set of systematically gathered grey and scientific literature. In conclusion, our taxonomy differs from the presented related work in the following ways. First, we provide factors of influence on several abstraction levels (i.e., themes, categories, and factors). Second, we do not focus on any domain, thus ensuring the taxonomy is valuable to researchers and practitioners of all fields.

\paragraph{Limitations}
Although the taxonomy was constructed in a way to mitigate threats of validity (e.g., several researchers worked on the literature review and data synthesis), the final taxonomy has several limitations. First, we took a software and data engineering perspective, thus not considering managerial or business-related factors. Second, although different types of data were considered when analyzing the literature, there is more evidence of challenges and data quality for tabular data than for other data types. This aspect can lead to a biased representation of IFs for different data types. Third, it is important to bear in mind possible IF relationships. In a study on factors influencing software development productivity, Trendowicz \& Muench \cite{Trendowicz&Muench2009} described dependencies between these factors. Similarly, data quality IFs may also be in a casual relationship which determines the final change of the quality of the data delivered by a pipeline. Nevertheless, more research is needed to investigate these relationships between different factors and examine their interdependence. 

%--------------------------------------------------------------------
%  Data-related and data pipeline processing issues
%--------------------------------------------------------------------

\subsection{Data-related and data pipeline processing issues}

\label{subsec:empirical_studies}
By mining GitHub and Stack Overflow, we got more profound insights into the root causes of data-related issues, their location in data pipelines, and the main topics of data pipeline processing issues for developers.

\paragraph{Interpretation} The majority of the analyzed issues on GitHub were caused by incorrect data types. A possible explanation for this may be that many data handling and ML libraries use custom data types which often cause interoperability issues (i.e., type mismatch) within pipelines \cite{Islam_etal2019,Islam_etal2019b,Zhang_etal2020}. This matches with the results of analyzing the main data pipeline processing topics developers ask on Stack Overflow. In fact, compatibility issues were found to represent a separate topic besides the typical data processing stages in pipelines. Regarding these stages, we found data integration and ingestion to be the most asked topics. This is in good agreement with the work of Kandel et al. \cite{Kandel_etal2012}, who found integrating and ingesting data to be the most difficult tasks for data analysts. A further aspect worth mentioning is that data-related issues rarely occur in the data transformation stage of a pipeline. A possible explanation for this is that incorrect transformations do not cause errors that are immediately recognized and thus can stay undetected. A further interesting finding was that data-related issues mainly occurred in the data cleaning stage of a pipeline. In contrast, developers ask the most about integrating, ingesting, and loading data but not about cleaning data. Although we cannot reason that the amount of asked questions of a topic reflects its difficulty, this can partly be explained by the fact that most issues in the data cleaning stage can be attributed to the raw data and data type characteristics and, thus, not directly to developers' skills. 

\paragraph{Comparison to related work}
We first compare our case study on GitHub with the work of Rahman \& Farhana \cite{Rahman&Farhana2021}. In their work, the authors investigated the bug occurrence in Covid-19 software projects on GitHub. They found data-related bugs to be their own category and identified storage, mining, location, and time series as corresponding sub-categories. However, their classification of data-related issues is strongly based on the inherent characteristics of Covid-19 software. For example, location data bugs describe issues where geographical location information in the data is incorrect. In contrast, our work describes general root causes applicable to a broader range of domains. A further study that dealt with data bugs was published by Islam et al. \cite{Islam_etal2019b}. In their paper, the authors identified data-related issues as the most occurring bug type in deep learning software. However, the authors do not provide further empirical evidence on concrete root causes of these issues. 

Regarding mining Stack Overflow, several other contributions used this platform to investigate topics and challenges for developers in related fields. Prior work \cite{Alshangiti_etal2019,Islam_etal2019} analyzed difficulties for developers in the closely related area of ML pipelines. Both studies found data preprocessing to be the second most challenging stage. However, these studies focused in particular on preparing data for ML models, e.g., data labeling and feature extraction. A further study \cite{Wang_etal2022} on issues encountered by Apache Spark developers identified data processing as the most prevalent issue, accounting for 43\% of all questions asked. However, this study did not detail the main topics of data processing and maps them to the usual processing stage of a pipeline.

\paragraph{Limitations}
To maintain the internal validity of the empirical studies, two researchers worked on the labeling independently. Inconsistencies were discussed and resolved together. However, the conducted exploratory case studies have several limitations. First, we solely used GitHub and Stack Overflow for our study. Thus, the generalizability of our results must be treated with caution. Second, the findings are limited by the tags used in both studies. To reduce these threats, we applied the snowballing method in defining the Stack Overflow tags and, besides using data-related tags in mining GitHub, non-data-related tags to ensure excluding only non-relevant issues. The findings of the studies are further limited by the fact that we relied on sampling during the analysis. Thus, we cannot guarantee the completeness of the identified results, although we chose a statistically significant sample of posts and issues for the detailed analysis. Moreover, similarly to the taxonomy limitations, most Stack Overflow posts are related to tabular data; thus the results can be biased towards other data types.

%--------------------------------------------------------------------
%  Overview of results and their relation
%--------------------------------------------------------------------

\subsection{Overview of results and their relation} \label{summary}
\begin{figure}[!h]
\centering
 \includegraphics[width=0.8\linewidth,keepaspectratio]{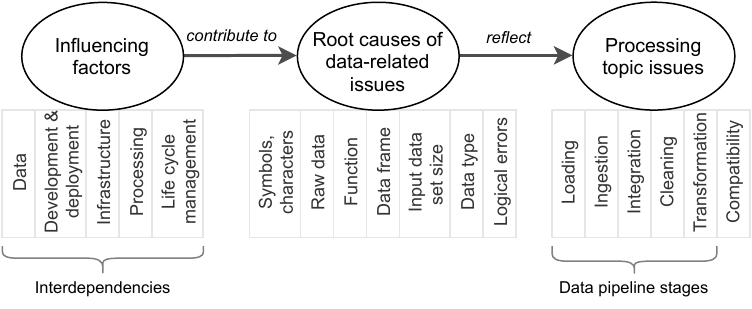}
 \caption{Overview of influencing factors' contribution to data-related issues and processing topic issues in data pipelines
 \label{fig:summary}
}
\end{figure}

Recapitulating our main research objectives, we recognize the following relations shown in Figure \ref{fig:summary}. Influencing factors contribute to the occurrence of the root causes of data-related issues. For example, the influencing factors 'data type handling' and 'data representation' explain the most frequent root cause 'data type'. This, in turn, suggests an interdependence between the influencing factors. Further, the typical data pipeline stages where data-related issues occur mainly reflect what developers ask about data pipeline processing. However, we found compatibility issues as a separate category of questions developers ask. It is possible that this category reflects the most identified root cause data type, especially as previous research links these aspects \cite{Islam_etal2019,Islam_etal2019b,Zhang_etal2020}. 

%% file: sections/7-Conclusions.tex
\section{Conclusions} \label{sec:conclusion}
This article contributes to enhancing the understanding of data pipeline quality. First, we descriptively summarized relevant data pipeline influencing factors in the form of a taxonomy. The developed taxonomy of influencing factors emphasizes today's complexity of data pipelines by describing 41 influencing factors. The influencing ability of the proposed factors was confirmed by expert interviews. Second, we explored the quality of data pipelines from a technological perspective. Therefore, we conducted an empirical study based on GitHub issues and Stack Overflow posts. Mining GitHub issues revealed that most data-related issues occur in the data cleaning (35\%) stage and are typically caused by data type problems. Studying Stack Overflow questions showed that data integration and ingestion are the most frequently asked topic issues of developers (47\%). In addition, we further found that compatibility issues are a separate problem area developers face alongside the usual issues in the phases of data pipeline processing (i.e., data loading, ingestion, integration, cleaning, and transformation). 

The results of our research have several practical implications. First, practitioners can use the taxonomy to identify aspects that may negatively affect their data pipeline products. Thus, the taxonomy can serve as a clear framework to assess, analyze, and improve the quality of their data pipelines. Second, the root causes of data-related issues and their specific locations within data pipelines can help industry practitioners prioritize their efforts in addressing the most critical points of failure and enhancing the overall reliability of their data products. Third, the main data processing topics of concern for developers enable companies to focus on common challenges faced by developers, particularly in data integration and ingestion tasks, which can lead to more effective support and assistance for developers in tackling those issues.

Moreover, this study lays the groundwork for future research into data pipeline quality. First, further research should be carried out to explore the identified compatibility and data type issues developers face during engineering pipelines in more detail. Second, future studies need to determine a ranking of the IFs based on the analysis of the individual influence levels of each factor. Third, further work may extend our study aiming to infer the difficulty of each data processing stage by additionally considering difficulty metrics of Stack Overflow posts, e.g., the average time to receive an accepted answer or the average number of answers and views.

We intend to focus our future research on studying the dependencies between the proposed influencing factors. For this purpose, we plan to apply interpretive structural modeling. There are already promising applications of this modeling technique to uncover interrelationships between factors \cite{Samantra_etal2016}.

%presented a taxonomy of factors influencing data pipelines' ability to deliver quality data. The taxonomy comprises 41 influencing factors (IFs) grouped into 14 IF categories covering five main topics: data, development and deployment, infrastructure, life cycle management, and processing. To validate the taxonomy, we conducted six structured interviews with experts who confirmed the proposed IF categories to be influential. 

%Third, we investigated the main data processing topic issues for developers in data pipelines by mining Stack Overflow posts and compared the identified topics with the typical data pipeline processing stages. 

%Second, we investigated the root causes of data-related issues and their location in data pipelines by analyzing issues from GitHub projects. 